\documentclass[namedreferences]{SolarPhysics}
\usepackage{epsfig}

\newcommand{\be}{\begin{equation}}
\newcommand{\ee}{\end{equation}}
\newcommand{\beq}{\begin{eqnarray}}
\newcommand{\eeq}{\end{eqnarray}}

\def\dj{d\kern-0.4em\char"16\kern-0.1em}

\begin{document}
\begin{article}
\begin{opening}

\title{ The Energy of High Frequency Waves in the Low Solar Chromosphere}

\runningtitle{ Energy of high frequency waves in the low solar chromosphere}    

\author{ A.\surname{An\dj i\'c}$^{1,2}$}
\runningauthor{ A.~An\dj i\'c} 

\institute{$^{1}$ Astronomical Research Centre, Queen's University Belfast, University Road, Belfast, BT7 1NN, UK 
                  \email{a.andic@qub.ac.uk}\\ 
	  $^{2}$ Institut f\"ur Astrophysik, Friedrich-Hund-Platz 1, 37077 G\"ottingen, Germany\\
             }

\date{Received: date}
% The correct date will be entered by the editor

%\maketitle

\begin{abstract}
High frequency acoustic waves have been suggested as a source of mechanical heating in the chromosphere. In this work the radial component of waves in the frequency interval $22$mHz to $1$mHz are investigated. Observations were performed using 2D spectroscopy in the spectral lines of Fe {\sc i} $543.45$nm and Fe {\sc i} $543.29$nm  at the Vacuum Tower Telescope, Tenerife, Spain. Speckle reconstruction has been applied to the observations.  We have used Fourier and wavelet techniques to identify oscillatory power. The energy flux  is estimated assuming that all observed oscillations are acoustics running waves. We find that the estimated energy flux is not sufficient to cover the chromospheric radiative losses.

%\keywords{chromosphere \and oscillation \and heating}
\end{abstract}
\keywords{Sun: chromosphere, Sun: oscillations, Sun: heating}
\end{opening}

\section{Introduction}
\label{intro}

The temperature of the Solar atmosphere first  decreases with height, reaching a minimum at approximately $500$km \footnote{where the region with $\tau_{500}=1$ is taken as the  reference point.} and then increases until it reaches coronal values \cite{vernazza81}. During the last century, high frequency acoustic waves were suggested as a source of mechanical heating. Their origin was believed to be below the photosphere. Propagating upwards through the solar atmosphere, they form shocks and dissipate energy in the chromosphere.  Ulmschneider (1971a, 1971b, 2003) suggested that waves with frequencies from $6$mHz to $100$mHz are the main carrier of the required energy and the peak energy transport should occur with waves of frequencies above $20$mHz.\par
 An important question of the chromospheric heating problem is whether mechanical heating is sufficient to balance radiative losses ({\it e.g.}, Kalkofen, 2001).   The energy required to balance the radiative losses just above the temperature minimum is around $14 \times 10^3 \,$ Wm$^{-2}$ \cite{athay89}. The existence of the chromosphere depends on a constant energy supply, in time, provided by mechanical heating \cite{ulmschneider03}.\par
The generation of acoustic waves can be described by the `Lighthill mechanism' \cite{lighthill51}. Acoustic waves appear to be generated by turbulence eddies for which the dissipation of kinetic energy by viscosity is negligible \cite{proudman52}. The majority of acoustic waves originate from isolated sources which occupy relatively small volumes, but the nature of those sources is not clear \cite{deubner83}. The most energetic oscillations should be generated in those regions where the convective velocities are largest.  The energy carried by the waves is sensitive to the Mach number of the turbulence and to the frequency of the emitted waves \cite{stein67}.\par
High frequency acoustic waves will form shocks within a few scale heights; low frequency waves need larger scales to shock \cite{stein74}. Heating by dissipation of acoustic shocks is time dependent \cite{carlsson95}. \par
 During wave propagation through the solar atmosphere their velocity amplitude increases with decreasing density, which will cause a height-dependent variation of their frequency. These changes are caused by the resonance absorption, the merging of shocks, and from shocks `cannibalizing' each other \cite{ulmschneider03}. Cram, Keil and Ulmschneider (1979) and Durrant (1980) recommend the use of the observed velocity fields and the analysis of the micro turbulent velocities caused by high frequency acoustic waves.\par 
Several observations detected high frequency waves (Liu, 1974; Mein and Mein, 1976; von Uexk\"ull {\it et al.}, 1985; Hansteen, Betta and Carlsson, 2000; Wunnenberg, Andjic and Kneer, 2003). The observed energy has been discussed by several authors (Liu, 1974; Lites and Chipman, 1979 ) who concluded that it is too small to cover the chromospheric losses. However, from the theoretical standpoint, there is a possibility that acoustic waves carry enough energy flux to cover the needs of the chromosphere (Umschneider, 2003; Kalkofen, 1990, 2001).  Wunnenberg, Kneer and Hirzberger (2002) claim that short period acoustic waves have enough energy for heating the chromosphere.\par
  In the recent works of Dom\`\i nguez (2004), Andjic and Wiehr (2006) and Andic and Vo\'cki\'c (2007) there is evidence for magnetic activity even in the so called ``non- magnetic" internetwork. This would question the assumption that the observed high oscillations are purely acoustic in origin. Based on TRACE observations Fossum and Carlsson (2006) claim there is not enough acoustic energy flux to heat the chromosphere and that magnetic fields in the internetwork are more significant than previously thought. 

\section{Observations}
\label{observations}

The data sets used in the analysis of this work were obtained using high-cadence observations of the Fe {\sc i} spectral lines  $543.45$nm ($g_{\rm L}=0$) and $543.29$nm ($g_{L}=0.335$). The 2D spectroscopy was performed using the German Vacuum Tower telescope (VTT) on the Canary islands, with the ``G\"ottingen" Fabry-Perot spectrometer \cite{bendlin92}. The instrument setup used in this work has been described in detail by Koschinsky, Kneer and Hirzberger (2001). We have not used a polarimeter in this study.  \par 
The post focus instrument is made following the demands of  speckle reconstruction (Bendlin, Volkmer and Kneer, 1992; Koschinsky, Kneer and Hirzberger, 2001). Therefore it takes simultaneously two types of images, broadband and narrow-band. The broadband images are images of integrated light in the observed spectral line. The narrow band images are scans through the spectral line using the Fabry-Perot spectrometer. For the scanning, a number of positions in the line,  a number of images per position and a exposure time were determined at the beginning of each observation taking into account the atmospheric conditions and requirements imposed by speckle reconstruction (Section \ref{data}). In atmospheric conditions with good or very good seeing, the cadence was higher, while during average seeing the cadence was lower. For the data sets where only one line was used, the line is scanned using $13$ positions across the profile, of which $2$ positions were in the continuum. In the data set with the two lines $18$ positions were used, $10$ for the stronger line ($543.45$nm) and $8$ for the weaker line ($543.29$nm) and only one position in the continuum.

\begin{table}
\caption{ Information about the data sets used in this work, dates, used lines and object of observation for obtained data sets.  The areas marked as `Quiet Sun' are quiet internetwork. }
\label{tpodaci}       
\begin{tabular}{lllllll}
\hline\noalign{\smallskip}
Data & Lines used & Object of & Coordinates& Images& Exposure & Cadence\\ 
set&&observation&&&[ms]&[s]\\
\noalign{\smallskip}\hline\noalign{\smallskip}
DS1 &543.45 and 543.29& Quiet Sun&$[0,0]$ & 108 & 30 & 28.4\\
DS2 &543.45 and 543.29& Bright points&$[0,0]$ & 108 & 30 & 28.4\\
DS3 & 543.45 & Pore& $[0,0]$ &  91 & 20 & 22.7\\ 
DS4 & 543.45 & Quiet Sun& $[96.7,90.7]$  & 91 & 20 & 22.7\\ 
\noalign{\smallskip}\hline
\end{tabular}
\end{table}

 The data sets were taken during an observing run in June 2004 (Table \ref{tpodaci}). The adaptive optics system was used (Berkefeld, Soltau, von der L\"uhe, 2003; Soltau {\it et al.}, 2002). \par
One of our main objectives was to achieve the highest cadence possible, which allows us to study waves with the highest frequencies. Since the time evolution of high frequency waves was of main interest, reasonably long time sequences were taken. Solar areas, which showed neither G-band structures nor other magnetic activity, were  classified as quiet internetwork regions. For active areas, we selected fields of view with the pores and large numbers of G-band bright points. The coordinates (Table \ref{tpodaci}) represent the position of the field of view, in arc sec, where the reference point is the centre of the Sun. \par

\section{Data Reduction and Analysis Methods}
\label{data}

The data reduction for the broadband images was done with a program developed by P. Sutterlin following the method of de Boer, Kneer and Nesis (1992) and de Boer and Kneer (1994). The program included the following procedures: subtraction of darks, flat fielding and speckle reconstruction. The program used for the reconstruction of narrow band images was developed by Janssen (2003) and is based on the method developed by Keller and von der L\"uhe (1992). \par
 The line bisectors  were calculated for each line profile in the 2D spectra to obtain more precise Doppler velocities, since the thermal drifts of the spectrometer may cause the first wavelength position to vary in each scanned line profile. The calculation was performed for each pixel in the reconstructed narrow-band images. All images from one scan are correlated, producing a  line profile for each spatial position. The line profiles were then interpolated by a cubic `spline' function and the bisectors calculated. Interpolation is necessary since the spectral line is scanned with a small number of positions in order to maximize the cadence.  As a reference point for the calculation of Doppler velocities we took a mean Doppler shift from one 2D spectrum and set it to zero. The procedure is repeated for all data sets. The shift of the bisector value for the core of the line with respect to this reference point corresponds to a Doppler shift, since the wavelength distance between the two scanning positions is known.\par
 Velocity maps used in this work are obtained from the bisector level corresponding to the core of the line profile.

\begin{figure}    %%%%%%%%%%%%%%%%%% FIGURE 1   
\begin{center}
\hbox{
\psfig{file=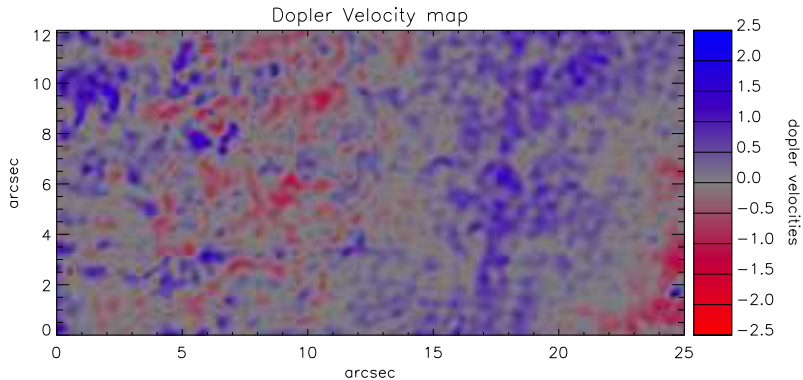,width=0.9\textwidth,clip=}}
\caption{Example of a velocity map at a height of the line core, from the data set DS4.}
\label{mapa200}
\end{center}  
\end{figure}

The time evolution was studied by converting the velocity maps into data cubes with time as the third axis. These maps have different sizes as a consequence of the reconstruction procedure of the narrow-band images. It is therefore necessary to do an additional correlation and alignment. This procedure can be carried out accurately only for images where clear structures are visible. Therefore it is done from the reconstructed broadband images which were already cut to the same dimensions as the narrow-band images.  The parameters obtained in this way were used for the alignment of the velocity measurements.\par
\begin{figure}    %%%%%%%%%%%%%%%%%% FIGURE 1   
\begin{center}
\hbox{
\psfig{file=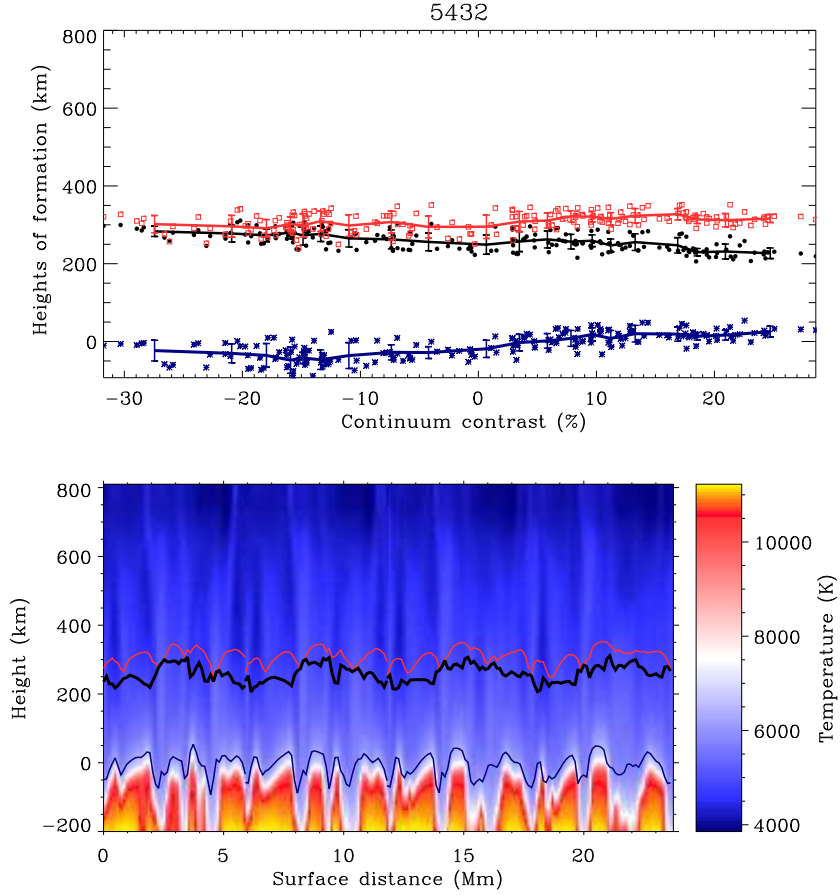,width=0.9\textwidth,clip=}}
\caption{Example of a results for the height of the line formation calculated for the spectral line Fe {\sc i} $453.29$nm. The top panel shows the heights vs continuum contrast. The $y$-axis gives height of formation in [km]. The LTE case is red color the NLTE is in the black for the line centre. The thin blue line  is the height of formation of the line wing {\it e.g.} continuum. The bottom panel shows the heights vs position along the "solar surface" in Mm. The background shows the vertical temperature of the model. The granules are visible as bright yellow and red features. (With courtesy of the N. Shchukina)}
\label{linije}
\end{center}  
\end{figure}

In order to determine the height levels corresponding to the calculated bisector levels, the formation layer of the line core have to be known.  Our estimate for the formation height of the Fe {\sc i} 543.45 and 543. 29 line cores is based on the location of optical depth unity at the central wavelength as calculated from a 3-D radiative-hydrodynamic simulation of Asplund {\it et al.} (2000). An example of the results obtained is shown at the Figure \ref{linije}. This method is described in detail in the work of Shchukina and Trujillo Bueno (2001).

 The average formation heights in LTE, are $308.2$km and $660.2$km for Fe {\sc i} $543.29$nm and Fe {\sc i} $543.45$nm respectively  (N. Shchukina, private correspondence). These formation heights are very close to those obtained with the LTE model of Holweger and M\"uller (1974). \par
The solar signal exhibits non- stationarity in both the spatial and temporal domains. Wavelet analysis is, therefore,  better suited than Fourier analysis for the interpretation of the observations. For our analysis we chose the Morlet wavelet:

\begin{equation}
\psi_0(t)= \pi^{-\frac{1}{4}}e^{i \omega_0 t} e^{-\frac{t^2}{2}},
\label{vaveleti1}
\end{equation}

\noindent where $\omega_0$ is the non-dimensional frequency and $t$ the non-dimensional time parameter. This wavelet is non-orthogonal, based on a Gaussian function and sinusoidal wave. The wavelet analysis code used here is based on the work by Torrence and Compo (1998).  The associated Fourier period, $P$, is $1.03s$ for $\omega_0 = 6$, where $s$ is the wavelet scale (see Table 1 in Torrence and Compo (1998)).  The wavelet transform is a convolution of the time series with the analyzing wavelet, thus a complete wavelet transform is achieved by varying the wavelet scale, which controls both the period and temporal extent of the function and scanning this through the time series. One-dimensional wavelet analysis is done for each spatial coordinate. At the beginning and end of the wavelet transforms there are regions where spurious power may arise as a result of the finite extent of the time series.  These regions are  referred to as the cone of influence (COI), having a temporal extent equal to the $e$-folding time ($t_d$) of the wavelet function. In our case it is:

\begin{equation}
t_d= \sqrt{2s}=\sqrt{2}\frac{P}{1.03}.
\label{vaveleti2}
\end{equation}

This time scale is the response of the wavelet function to noise spikes and is used in our detection criteria by requiring that reliable oscillations have a duration greater than $t_d$ outside the COI. This imposed a maximum period of $912$s (or $\approx 1$mHz) above which any detected periods were disregarded.  The automated wavelet analysis carried out in this work has previously been presented in detail by Bloomfield {\it et al.} (2006). In the current study only the power outside the COI is retained.\par
To estimate the energy flux we made the assumption that all registered oscillations are acoustic. Therefore, the energy flux is estimated using the expression:

\begin{equation}
F_h= \rho_h v^2 c_h,
\label{fluks}
\end{equation}

\noindent where, $\rho_h$ is the density at the height $h$, and $c_h$ the corresponding speed of sound. The values of $\rho_h$ and $c_h$ used in this work are presented in Table \ref{vrjednosti}. 

\begin{table}
\caption{Values of sound speed and density used for calculations of energy flux.} 
\label{vrjednosti}
%\begin{center}
\begin{tabular}{lll}
\hline\noalign{\smallskip}
Height [km]& Speed of sound [m s$^{-2}$] & Density [kg m$^{-3}$] \\[3pt]
\noalign{\smallskip}\hline\noalign{\smallskip}
$660$ & $6742$ & $1.32 \times 10^{-6}$ \\ 
$308$ & $7142$ & $3.46 \times 10^{-5}$\\ 
\noalign{\smallskip}\hline
\end{tabular}
%\end{center}
\end{table}

To determine the velocity of the particles, two methods were used.  Doppler velocities were obtained from the line bisectors with a method similar to the one used by Wunnenberg, Kneer and Hirzberger (2002) and they were used as representative of the velocity of the detected oscillations. This procedure is conducted over the whole data set regardless of the location of the detected oscillations. In the second method wavelet analysis was used. Wavelet analysis contains informations on both the amplitude and phase of the detected oscillations.  The velocity amplitudes determined by the wavelets are used in Equation (\ref{fluks}). The majority of the Doppler velocities shown in the velocity maps belong to non-oscillatory behavior. Only amplitudes which belong to the detected oscillations are taken into account here. 

\section{Results}
\label{results}

\begin{figure}    %%%%%%%%%%%%%%%%%% FIGURE 2   
\begin{center}
\hbox{
\psfig{file=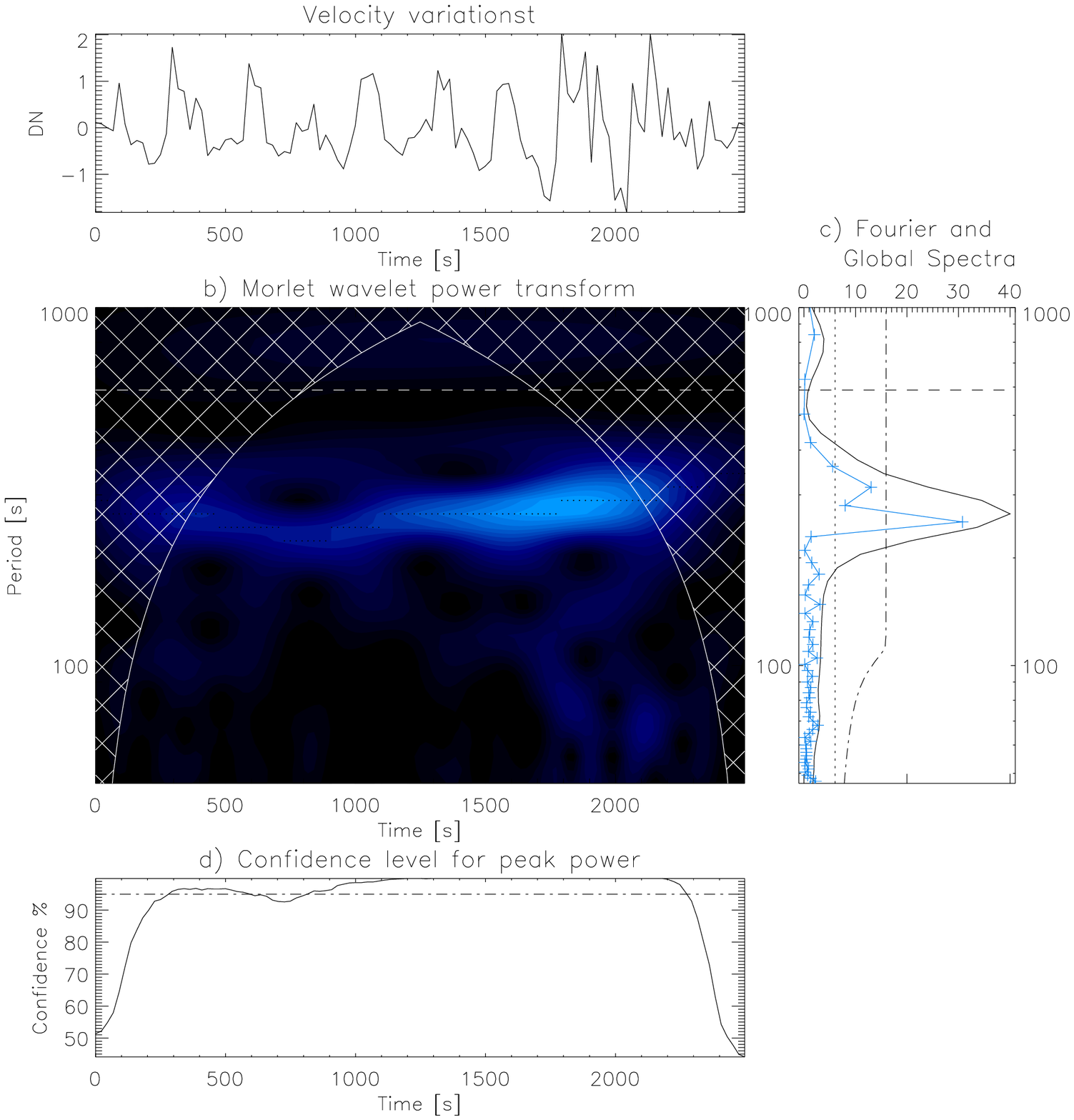,width=0.5\textwidth,clip=}
\psfig{file=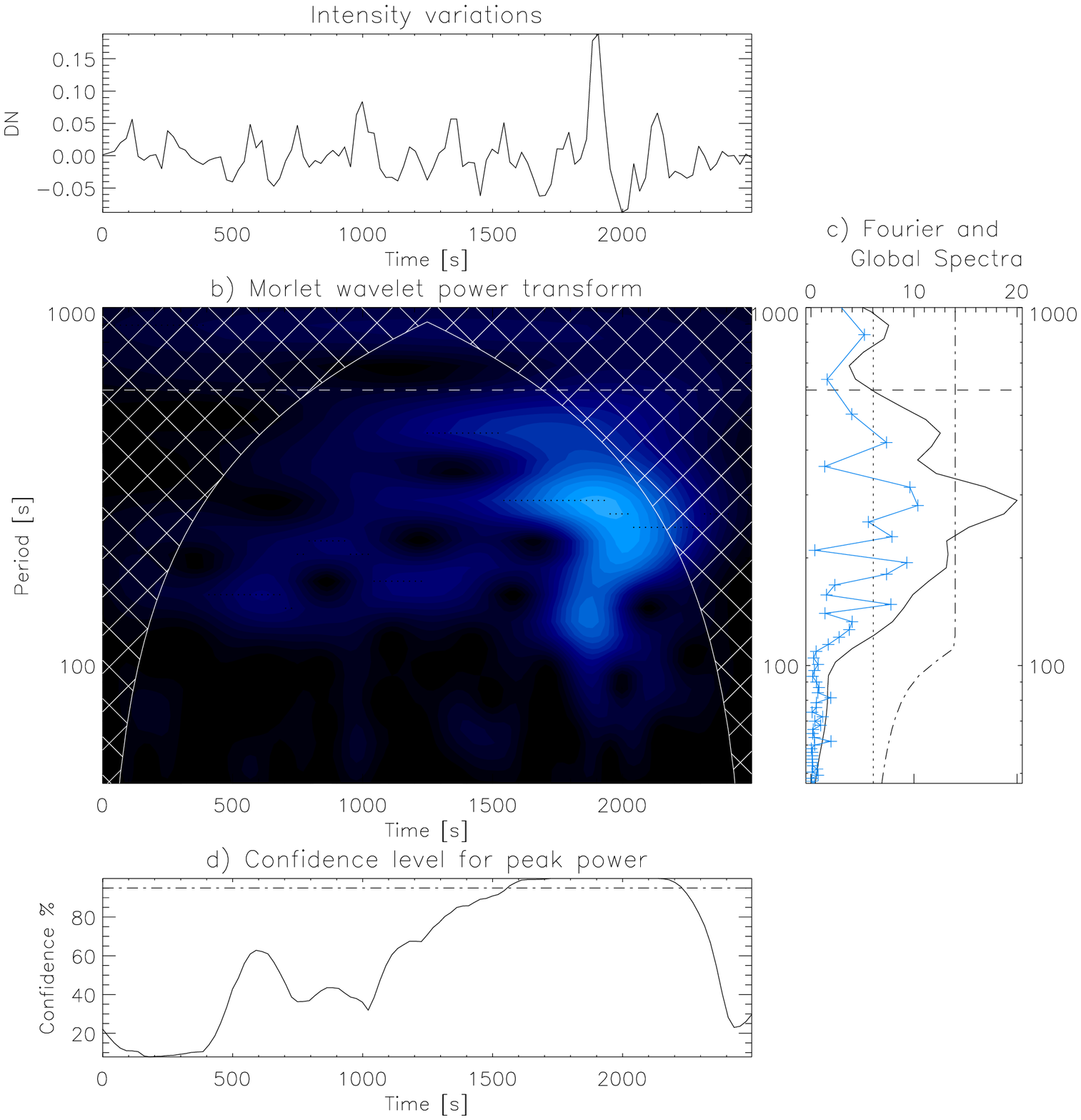,width=0.5\textwidth,clip=}}
\caption{The oscillations with the largest amplitudes within the frequency range of $22$mHz to $1$mHz. The left panel shows the oscillations observed within the Doppler velocity temporal variation and the right panel oscillations observed within the intensity variation. Top panels marked with Velocity and Intensity variations present the actual signal which is submitted to the wavelet analysis. Below the result of the wavelet analysis is showed, marked with Morlet wavelet power transform, the registered power is presented in the color-coded-multilevel contours, where the minimum power corresponds to the black color and the maximum to the white. The dashed line on this presentation represent the maximum acceptable period detection. On the bottom of each wavelet presentation the confidence level for the peak power is presented and on the right side global and Fourier spectra are plotted for each signal, dash-doted line in this diagram present the statistical significance of the registered power. }
\label{nfp9}
\end{center}  
\end{figure}

  Our data show a wide range of oscillations. Figure \ref{nfp9} shows the oscillations with the largest amplitudes within the frequency range of $1$mHz to $22$mHz.  The same range  of frequencies were registered in both velocity and intensity. After an inspection by eye of light curve and cycle duration of the wave trains no direct evidence for a correlation in the two waveforms is found.

The duration of the oscillations is also analyzed. This parameter can be useful in interpreting the energy associated with a wave \cite{bloomfield06}. The detections were sampled in terms of their duration with uniform full-cycle binning with a separate analysis been preformed in the velocity and intensity space.\par

\begin{figure}    %%%%%%%%%%%%%%%%%% FIGURE 2   
\begin{center}
\hbox{
\psfig{file=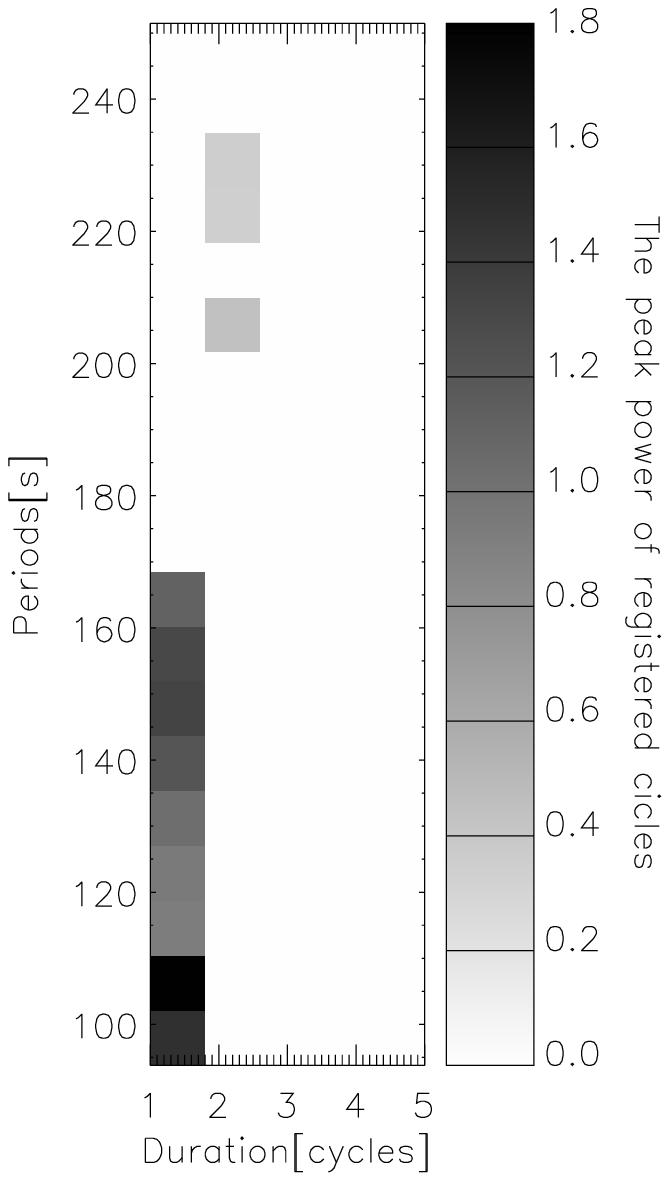,width=0.5\textwidth,clip=}
\psfig{file=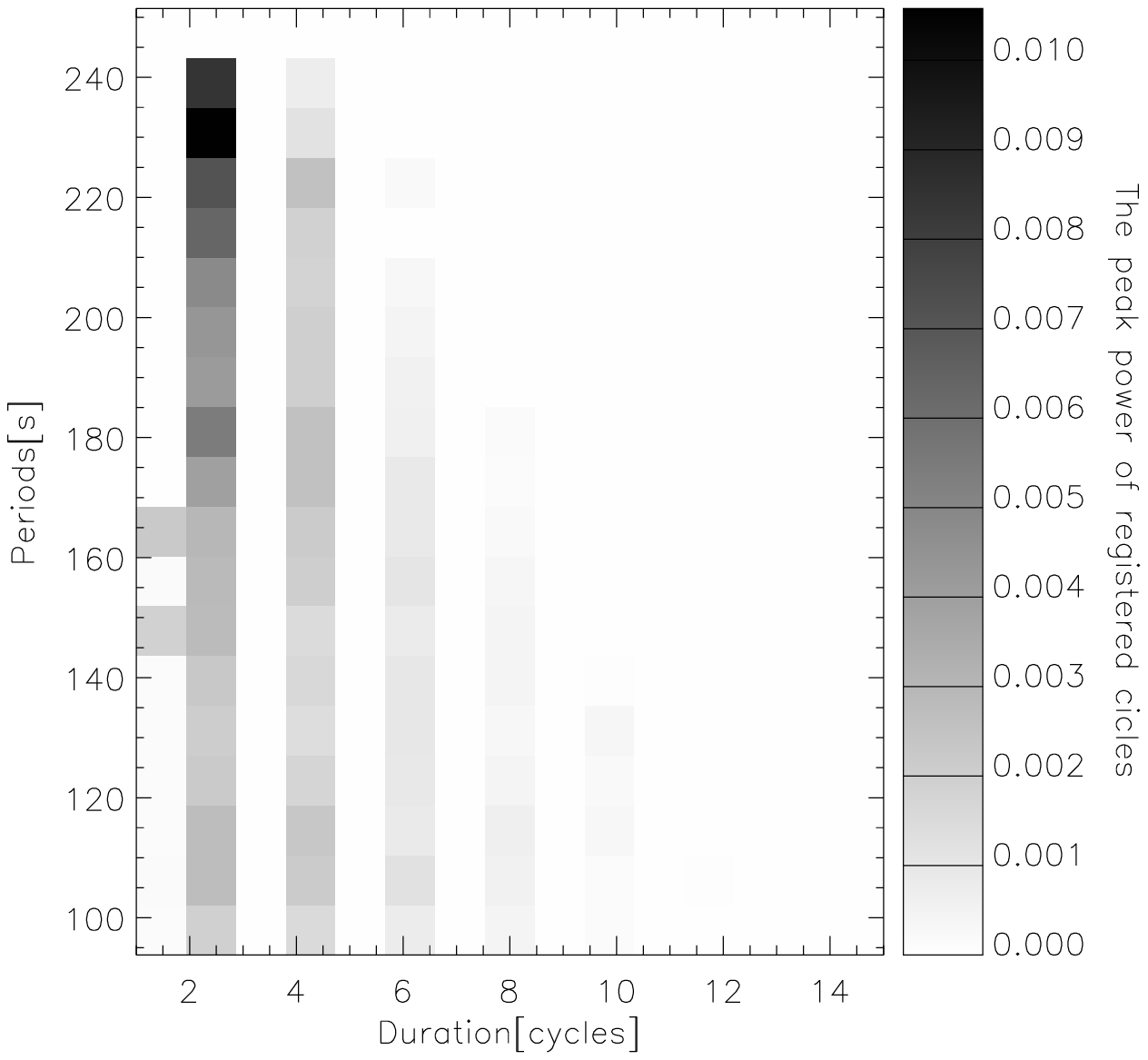,width=0.5\textwidth,clip=}}
\caption{Period distribution of observed velocity ( left) and intensity ( right) oscillations (ordinate) as a function of duration of wavelet detection (abscissa). The various shades of gray represent the maximum power of the observed oscillation cycle in accordance with the scale shown. This image presents non-filtered data for the core of the observed spectral line.}
\label{cik1}
\end{center}  
\end{figure}

Figure \ref{cik1} shows the period distribution of the observed velocity and intensity oscillations as a function of the number of cycles. It is noticeable that not the same wave trains are observed. Also it is noticeable that most of the power for the velocity oscillations is contained within waves with duration of one cycle, while intensity oscillations last for a larger number of cycles.

\subsection{Energy Transport and the Dissipation of Energy}\label{energicno}

\begin{figure}    %%%%%%%%%%%%%%%%%% FIGURE 1   
\begin{center}
\hbox{
\psfig{file=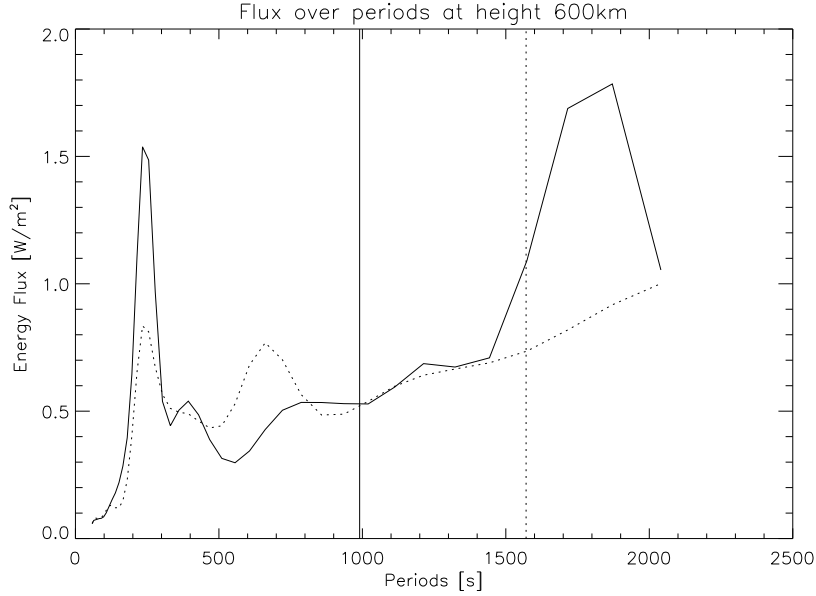,width=0.9\textwidth,clip=}}
\caption{Estimated acoustic flux using velocity maps ( ordinate) as a function of periods of registered oscillations (abscissa) at the height of $300$km. Solid lines represent results from the data set DS2. Dashed lines represent results from the data set DS1, while vertical lines represent the credible period.}
\label{energ1}
\end{center}  
\end{figure}

\begin{figure}    %%%%%%%%%%%%%%%%%% FIGURE 1   
\begin{center}
\hbox{
\psfig{file=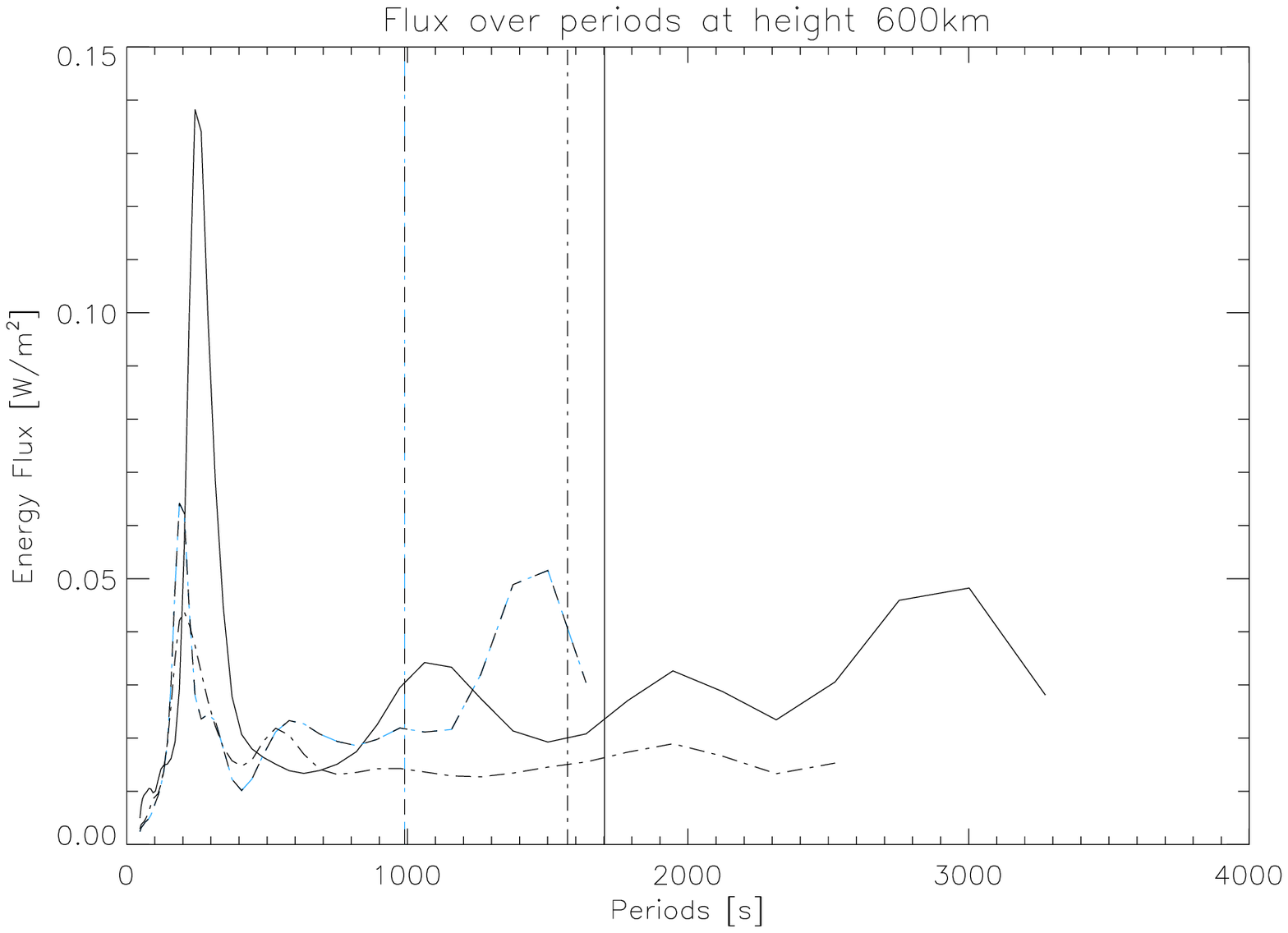,width=0.9\textwidth,clip=}}
\caption{Estimated acoustic flux using velocity maps (ordinate) as a function of periods of registered oscillations (abscissa) at the height of $660$km. Solid lines represent results from the data set DS4. Dashed lines represent results from the data set DS2,  blue dashed-doted lines results from the data set DS3 and dashed-doted line from the data set DS1. The vertical lines represent the credible period, where the style of the line correspond above assigned ones to the data sets.}
\label{energ3}
\end{center}  
\end{figure}

 To investigate how much energy is carried by acoustic waves the LTE model \cite{holweger74} was used to determine the density and corresponding speed of sound for heights used in the analysis.  We have made the assumption that all registered power is acoustic. The velocity profiles, taken from wavelet analysis, are inserted into Equation (\ref{fluks}), which provides an estimate of the energy flux. The calculated fluxes are presented in Figures \ref{energ1} and \ref{energ3}.

 An inspection of Figures \ref{energ1} and \ref{energ3} shows that the estimated flux is decreasing with period in both cases. The energy flux is larger at the lower atmospheric height.  The energy flux estimates derived from the velocity maps (Figures \ref{energ1} and \ref{energ3}) take into account that each pixel spends only a limited amount of time in oscillation. Therefore the average in this case is a reflection of the temporal behavior of the energy flux.\par
 Data set DS4 has a cadence of $22.7$s, making observations of high frequency waves possible. According to Ulmschneider (1971a, 1971b, 2003) there should be more energy flux detected when observations approach frequencies around $25$mHz. A higher cadence  should allows us to measure higher energy fluxes. Never the less, there is no significant increase in the estimated energy using DS4.\par
The first method for the estimating the particle velocity yields the results presented in Table\ref{fluksrezultat2}. There is no significant difference with the energy flux estimated using the bisector method of Wunnenberg, Kneer and Hirzberger (2002) and wavelet method.

\begin{table}
\caption{ Energy flux for the data sets DS6. The flux is calculated using the velocity averaged over the corresponding  velocity maps of the data set.}
\label{fluksrezultat2}     
%\centering  
\begin{tabular}{llllll}
\hline\noalign{\smallskip}
data set & cadence [s] & $F_{660}$ [W m$^{-2}$] & $F_{308}$ [W m$^{-2}$] & $v_{660}$ [W m$^{-2}$] & $v_{308}$ [km s$^{-1}$]\\[3pt]
\noalign{\smallskip}\hline\noalign{\smallskip}
DS1 & 28.4 &0.005 &0.103& 0.57& 0.42\\
DS2 & 28.4 &0.003 &0.064& 0.32& 0.26\\
DS3 & 28.3 &0.045 &N/A& 0.62& N/A\\
DS4 & 22.7 &0.005 &N/A& 0.62&N/A\\
\noalign{\smallskip}\hline
\end{tabular}
\end{table}

\section{Discussion and Conclusions}
\label{conclusions}

High frequency waves were detected at heights of $300$ and $660$km above the photosphere. The observations used in this work include internetwork, bright network with G-band structures and a pore. We find no evidence for a correlation between velocity and intensity oscillations. The estimated acoustic energy flux is found to be insufficient for heating the chromosphere. \par

\vskip4mm

  When establishing possible errors in observing high frequency waves, the whole observations and data reduction procedures need to be considered.  The influence of atmospherics seeing is mainly visible during speckle reconstruction. The work by von der L\"uhe (1984, 1993, and 1994) estimates an error of $10$\% in intensity introduced by speckle reconstruction. Of course, the most important part of reconstruction is the noise filtering. In this work we followed the recommendations of de Boer (1996), when setting the noise filter. \par
 The seeing influence is most visible with differing data cadences.  In the case of good seeing, cadence was lower (see Section \ref{observations}) than in the case of very good and excellent seeing conditions. Therefore the amount of noise remaining after speckle reconstruction tended to vary only inside predicted error levels (von der L\"uhe, 1984, 1993, 1994; de Boer, 1996).\par  
 To make an estimate of the observed energy flux several assumptions were made. The first assumption was the use of the LTE model \cite{holweger74}. Since the core of the line used is in the non-LTE area, being just above the temperature minimum \cite{vernazza81} and the strong lines of neutral iron do suffer from a non-LTE effect in both their opacity and in their source function \cite{shukina97}, future studies should include these effects.  \par

\vskip4mm
 Wunnenberg, Kneer and Hirzberger (2002) find no correlations between velocity and intensity oscillations and our findings agree with them. In our analysis the velocity oscillations tend to last for less cycles than the intensity oscillations.\par
 Previous studies on this topic have also found  a lack of sufficient energy flux from acoustic waves. The work by Schmieder and Mein (1980) found an energy flux of $2$ Wm$^{-2}$ over the frequency range $0$mHz to $10$mHz -the lower frequency range than the one presented here. The work by Schmieder and Mein (1980) was concentrated on the middle chromosphere, while here we study the upper photosphere and lower chromosphere. Also work by Lites and Chipman (1979) investigates the similar range of frequencies from $0$mHz to $11.1$mHz and state that the energy flux is too low to heat the chromosphere. The work by Mein and Mein (1976) observed high frequency waves in the range $10$mHz to $15$mHz at a height range of $800$ to $1800$km and also established that there is not enough energy present to heat the chromosphere. Liu (1974) observed the oscillations int he same frequency range and heights as the present work, but his estimated energy flux was  several  orders of magnitude higher than the energy fluxes found here. In light of the recent theoretical work (Kalkofen, 1990, 2001; Ulmschneider, 2003) the energy  estimated in the work of Liu (1974) should be enough for heating the chromosphere.  Wunnenberg, Kneer and Hirzberger (2002) reach the same conclusion. Our findings do not agree with them. One of the main differences between the work of Wunnenberg, Kneer and Hirzberger (2002) and the present study has to do with velocity response functions of the  Fe {\sc i} $543.45$nm line. In this work, the velocity response functions were not used and only oscillations which are $99$\% significant were taken into account.
 The recent study by Fossum and Carlsson (2006) has also concluded that acoustic waves do not carry sufficient energy flux for the heating of the chromosphere.\par

 The energy estimates on Figures \ref{energ1} and \ref{energ3} show a difference of one order of magnitude for the energy flux calculated for the spectral lines Fe {\sc i} $543.45$nm and Fe {\sc i} $543.29$nm. To get an estimate which reflect the temporal behavior for the corresponding data set only pixels which actually belong to the observed oscillation were taken into account. Since the majority of the observed oscillations were short lived we do not distinguish between waves of different duration. If we average over the whole time sequence, the observed oscillations cover $99$\% to $100$\% of the field of view for Fe {\sc i} $543.45$nm in DS1,DS2 and DS4. The corresponding coverage for Fe {\sc i} $543.29$nm is $72$\% to $81$\% in DS1 and DS2. Therefore the averaging for the energy estimates on Figures \ref{energ1} and \ref{energ3} is done for whole field of view. \par

One of the main assumptions used in the current and previous studies is that all registered oscillatory power is acoustic. This assumption is incorrect even for quiet internetwork regions. Dom\`\i nguez (2003) found that the magnetic flux in the quiet internetwork is smeared by insufficient resolution and therefore underestimated. He also found that there are areas in the quiet internetwork which have magnetic fields comparable to those from the pores. Trujillo Bueno, Shuckina and Asensio Ramos (2004) claim that we can only see $1$\% of the magnetic field present in quiet internetwork regions. Andjic and Wiehr (2006) and Andic and Vo\'cki\'c (2007) show evidence for some magnetic activity in quiet internetwork regions used in this work.

\begin{acknowledgements}
 I acknowledge the Ph.D. scholarship from Max Planck Institut for Solar System Research (Max Planck Institute f\"ur Sonnensystemforschung), Katlenburg-Lindau, Germany.\par 
I wish, especially, to thank Dr. N. Shchukina for calculating the formation heights for the used lines.\par
During the work itself, I had lots of stimulating discussions for which I wish to thank  R. Cameron, A. V\"ogler and E. Wiehr. I wish to thank P. Sutterlin and K. Janssen for giving me the software for the data reduction. For the help with the observations I wish to thank J. Hirzberger and K. Puschmann.\par
I also wish to thank M. Mathioudakis with help in formulating the sentences and together with D.B. Jess for help with the English. 

\end{acknowledgements}

\end{article} 
\end{document}